# Exploration of reproducibility issues in scientometric research
# Part 2: Conceptual reproducibility

Theresa Velden[*], Sybille Hinze[**], Andrea Scharnhorst[***], Jesper Wiborg Schneider[****], Ludo Waltman[*****]

[*] *velden@ztg.tu-berlin.de*
Zentrum für Technik und Gesellschaft (ZTG), Technische Universität Berlin, Berlin (Germany)

[**] *hinze@dzhw.eu*
Deutsches Zentrum für Hochschul- und Wissenschaftsforschung (DZHW), Berlin (Germany)

[***] *andrea.scharnhorst@dans.knaw.nl*
Data Archiving and Networked Services (DANS), Royal Netherlands Academy of Arts and Sciences, The Hague (the Netherlands)

[****] *jws@ps.au.dk*
Danish Centre for Studies in Research and Research Policy, Aarhus University, Aarhus (Denmark)

[*****] *waltmanlr@cwts.leidenuniv.nl*
Centre for Science and Technology Studies (CWTS), Leiden University, Leiden (the Netherlands)

**Introduction**

Several scientific fields are experiencing an intensive debate about the reliability of published results that centers on various aspects of reproducibility. In psychology and biomedicine, the discussion has been triggered in part by isolated cases of fraud (data fabrication in social psychology as well as in human genomics), and in part by concerns about the appropriateness of widely used research methods and the overselling of results that undermine the robustness of knowledge claims (Flier 2017; Open Science Collaboration 2015; Pashler & Harris 2012).

To explore the question how concerns about the reproducibility of research may translate to the field of scientometrics, we initiated a workshop at ISSI 2017 in Wuhan.[1] At the workshop several speakers suggested, that given differences in research objects, methods, and study designs, scientific fields differ with regard to the type and pervasiveness of threats to the reproducibility of their published research.

In the run up of the STI 2018 conference, we decided to explore how an assessment of the specific challenges to the reproducibility of research in the field of scientometrics could be conducted based on a critical review of research published in the field. To this end we distinguish different categories of studies, and developed a taxonomy of threats to reproducibility that may be identified by a review of published papers. This paper is the second part of our report about this explorative study.[2] In part 1 of our explorative study (Waltman et al. 2018), we focus on direct reproducibility - that is the exercise of a third party repeating a published study using the same method, data, and procedures. In this paper, part 2 of our study, we focus on conceptual reproducibility - that is the exercise of a third party testing the robustness of knowledge claims by reproducing the original claims using different data, methods, and procedures.

---

[1] Workshop report available online at www.issi-society.org/blog/posts/2017/november/reproducible-scientometrics-research-open-data-code-and-education-issi-2017/.

**Background**

The concept of reproducibility can refer to various approaches to and purposes of reproducing (some aspect of) an original study. What variety of reproducibility is seen as most pertinent, seems to vary by scientific domain. This diversity of perspectives has led to a thorough confusion of terminology around reproducibility, including antithetical definitions of the terms *replicability* and *reproducibility* (Goodman et al. 2016; Barba 2018). To cut through the thicket of terminological confusion, we use the term *reproducibility* as a generic umbrella term and focus on two distinct subtypes that we define below.

One way to think about theoretical differences between concepts of reproducibility is in terms of varying degrees of the similarity of conditions between the original study and a reproduction study, including the study design, methods, and data used (Chen 1994). We focus in our two-part exploratory study on the two subtypes that are located at opposite ends of this spectrum and have distinct scientific functions: *direct* and *conceptual* reproducibility (in line with Fidler et al. 2017).

*Direct reproducibility* is located at the 'greatest similarity' end of the spectrum where the same data, tools and methods are used to reproduce and verify a study with the expectation of obtaining the identical or very similar empirical result obtained in the original study, unless some unintended or fraudulent error was made in the original or the repetition study.

*Conceptual reproducibility* is located at the other end of the spectrum where a study is reproduced using different data, tools and methods with the aim of testing the robustness of the fundamental knowledge claim made by the original study. So instead of the specific evidence obtained in a study, it focuses on the interpretation of that evidence and the validity of the knowledge claims derived from that evidence.

While robustness of scientific knowledge is achieved only in a cumulative and discursive process within the scientific community and not by a single publication, we argue that individual publications provide the foundation for producing robust knowledge. Their contribution is twofold: first, through the accuracy of the empirical evidence that they generate, that is by delivering results that are directly reproducible; second, by articulating their knowledge claims in accordance with the empirical evidence they produce, that is by not using questionable research methods (see e.g. Simmons et al. 2011; Schneider 2015) and/or overstating claims.

---

[2] Both parts of our study were conducted using the same methodological approach, in particular the same categorization of study types, the same sample of publications, and a very similar taxonomy guided reviewing process. To allow each paper to be read independently, we opted to repeat those methodological descriptions rather than to refer the reader to the other paper.

Of the various causes of irreproducibility identified above, error, sloppiness, and fraud relate to the execution of the study and affect its direct reproducibility, whereas questionable research methods, overselling, and publication bias relate to the interpretation of the evidence and affect its conceptual reproducibility.

**Analytical approach**
To explore how one might identify reproducibility issues in publications of scientometric studies, we defined a categorization of types of scientometric studies and critically reviewed them with regard to potential threats to reproducibility (analogous to the data selection and procedure used in part 1 of our explorative study). To ensure consistency across our reviews, we developed a taxonomy of potential threats to conceptual reproducibility, presented further below.

*Classification of studies*
As a basis for our explorative study, we created a high-level classification of scientometric studies in order to explore how threats to reproducibility may vary by type of study. We distinguished theoretical/conceptual, methodological, and empirical studies, and further refined the empirical category in order to account for the large amount and variety of empirical studies in scientometrics. Our classification is presented in Table 1. As often in classification, many studies do not fit neatly into one of the five categories. We decided to assign papers to the categories based on the primary focus of a study.

Table 1: High-level classification of types of scientometric studies.

| Category no. | Name | Description |
| --- | --- | --- |
| 1 | Theoretical/Conceptual | Studies that are primarily theoretically/conceptually focused |
| 2 | Methods | Studies that are primarily methodologically focused. |
| 3 | Empirical (General) | Studies that are primarily empirically focused, aimed at answering substantive research questions in the study of science. |
| 4 | Empirical (Case) | Studies that are primarily empirically focused, taking a 'case study' approach, that is, focusing on analyzing particular research domains or particular countries, research institutions, or journals. These studies do not aim to develop more general insights that go beyond the particular case they analyze. |
| 5 | Empirical (Data Source) | Studies that are primarily empirically focused, aimed at getting a better understanding of the data sources available for scientometric research. |

*Taxonomy of threats to reproducibility*

The taxonomy for threats to conceptual reproducibility identifies substantive issues that undermine our confidence that the central knowledge claims made in a publication are robust, that is likely to hold up to a test by conceptual reproduction. Based on the debate in other fields that questionable research methods and overselling play a major role in explaining irreproducibility, we distinguishes between Operationalization assumptions and decisions, Quality Control, and Reporting of Results (see Figure 1). Within the category of Operationalization assumptions and decisions, we review whether the **selection of data**, **data modeling**, and the choice of the **analytical methods** is appropriate for the chosen research question. Within the category of Quality Control, the firmness of the research design is complemented by looking for evidence for measures for quality control. Among them, we looked into discussions on the **completeness, consistency of primary data**, and discussions how **parameter choices** influence the **stability of results**. Within the category of Reporting of Results, we looked if **limitations were explicitly stated**, **claims were backed-up by empirical results**, and if there was an **adequate discussion of limits in precision, measurement error, randomness.**

Figure 1: Taxonomy to identify potential threats to conceptual reproducibility of a published scientometric study.

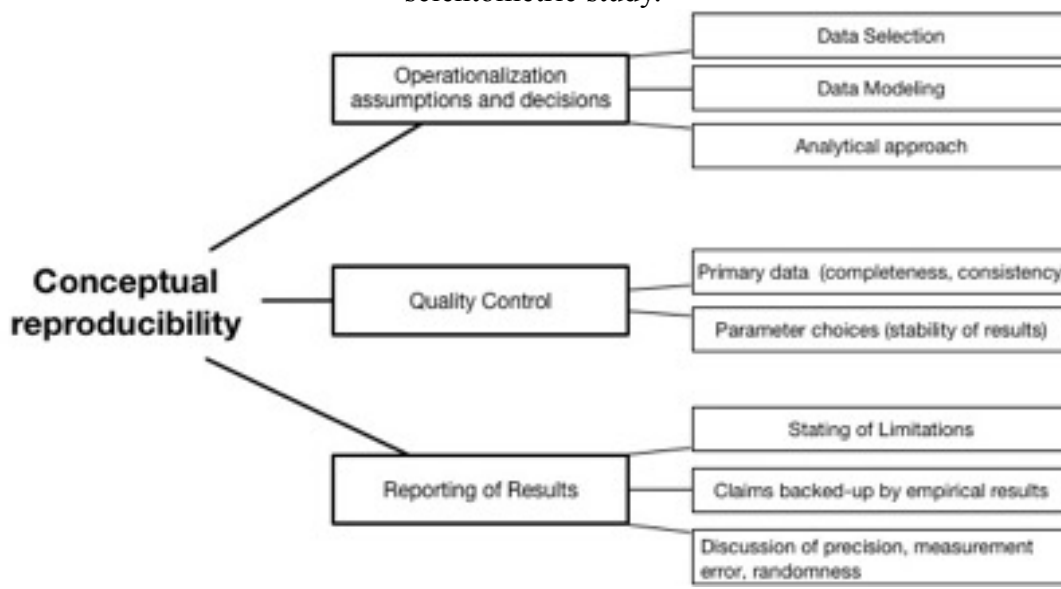

*Data and method*

We review the same set of papers that we selected for part 1 of our explorative study (Waltman et al. 2018) which contains one paper for each of the five study type categories, published within the last two years (two in *Scientometrics,* two in *Journal of Informetrics* and one was made available as a preprint in the arXiv). With the paper selection we aimed at selecting papers that serve as a good example of one the above five categories. Papers were selected and agreed upon unanimously by all authors of this paper. Each paper was then reviewed by at least two of the authors of this paper, one paper was reviewed by three. Each of the reviewers was asked to assess the papers regarding the elements identified by the taxonomy, to check for weaknesses that would lead the reviewer to suspect shortcomings in the robustness of the knowledge claims derived in the paper.

We do not reveal the identity of the five papers, but provide an overview of key features in Table 2. Our focus is on providing general insights into the reproducibility of scientometric research, not about the extent to which specific papers are reproducible. Readers who want to know more about the papers that were reviewed are invited to contact us.

Table 2: Properties of the five papers selected for review in this explorative study.

| Paper no. | Study type | Topic area | Methods | Data | Tools |
|---|---|---|---|---|---|
| 1 | Theoretical/ conceptual | Citation theory | Theoretical reasoning, simulation | Synthetic | Self-developed simulation software |
| 2 | Method development | Topic extraction | Network clustering | Bibliometric, proprietary, large-scale ($10^7$) | Open source software |
| 3 | Empirical (Substantive) | Innovation studies | Statistical regression analysis, network analysis | Patent data, proprietary | Standard, proprietary statistical package, network analysis tool (proprietary, free trial) |
| 4 | Empirical (Case) | Specialty study at national level | Network analysis and visualization | bibliometric, proprietary, small-scale ($10^3$) | Freely accessible online tool |
| 5 | Empirical (Data source) | Evaluation of sources for citation analysis | Recall and precision measurements, correlation coefficients | Bibliometric, proprietary and freely accessible large-scale ($10^5$-$10^6$) | Freely accessible online tool for query generation |

**Results**

Conceptual reproducibility focuses on the question whether knowledge claims published in a field are found to be robust when tested using an alternative approach with different data, methods, and study design. The scope of our assessment of the status of conceptual reproducibility in the field of scientometrics is very limited, as it is restricted to assessing the contribution that individual papers make through using research design that are appropriate to the research question being asked, and through formulating claims that are not overselling results but are supported by the evidence that the respective study has produced. However, what is seen as appropriate, related epistemic norms and values, are under constant debate and negotiation in a field, and therefore cannot be handled as a simple checklist. Consequently, scrutinizing the papers against those categories leaves more space for different judgment.

*Operationalization (assumptions, decisions)*
The question of data selection and modelling is obviously most relevant for *empirical studies*. Data selection and modelling should be consistent with the research problem a paper tackles. Reviewers did not always agree in their critical remarks. For instance, perspectives on how to delineate a field, or if the choice of a database is appropriate for a certain research question, vary within the scientometrics community. *Method papers* need to argue that the choice of data to demonstrate the value of their data analytic method is suitable to prove that claim. *Conceptual papers* can also contain data issues. In our example, the conceptual paper presented a toy data set and a simulation model - choices made for either can be challenged and gauged against empirical phenomena.

For a *method paper*, the subcategory choice of the analytical method is evidently the most central. We found differences in the extensiveness of how authors introduce into concepts and related methods The reviewers welcomed extensive discussion how to operationalise a certain research question; and if methods used were standard in the field. For the *empirical papers* though there was also critique about using standard tools without critical reflecting about limits of a tool.

To summarize, extensive discussions of the choice of data and methods were positively marked by the reviewers. In some cases, standard methods, tools and datasets were found to be taken too much for granted. A critical view on one's own approach and the articulation of pro and cons in the choices made relative to the specific research question pursued would instill greater confidence in the robustness of the results. The conceptual and method paper scored relatively high here, while the empirical papers in the eyes of the reviewers could have been more explicit or more critical.

*Quality control*
In this category we look for evidence for measures for quality control that could increase confidence in the robustness of results. For the *conceptual papers* this leads to the questions if choices made are thoroughly detailed. For a *method paper*, for instance the influence of noise in primary data on the methodological analysis can be an important point. For the *empirical papers* in our sample questions about the role of missing data, the exclusion of certain date from the analysis, and the representativeness of a certain method of data collection were posed. The *conceptual paper* and the *method paper* scored relatively well on those criteria, but the reviewers were more critical about the *empirical papers*. Either a discussion of completeness and consistency of data and the choice of parameters was entirely missing; or if present the consequences of such omissions for the argument of the article were not discussed.

*Reporting of results*
Positive is that all papers in our small sample addressed limitations of their studies, so there was clearly a self-critical attitude present. Remarks of authors on the limits to generalisability of results, the risk of obsolescence of the results when the data services used are changed, or the possibility to use another simulation model were usually appreciated by the reviewers. However, there were critical remarks concerning the extent to which specific claims were backed-up by the empirical results. In particular, *empirical papers* of category 4 (case study) seem to be susceptible to such an 'overplaying of your hands', especially when lacking details when discussing limits resulting from sample size. In the case of category 3 (substantive research question) critique on the reporting of limits and overstating of claims was voiced, mixed with doubts about the support the research method (in this case regression analysis) lent to the results.

**Discussion**

In our limited review of scientometrics publications, we found the technical preconditions for direct reproducibility in part 1 of our explorative study (Waltman et al. 2018) much easier to assess using a checklist approach than the likelihood of their conceptual reproducibility. We found that reviewing the articles for issues that may present a threat to their conceptual reproducibility largely mimicked the process and effort of conducting a typical peer-review of a journal article submitted for publication - significantly scaling down the number of publications we had hoped to review in each study type category in the time allotted to this explorative study - which in itself is one of the lessons learned.

The taxonomy directed our attention to specific aspects, such as the adequacy of study designs and methods, and the adequacy of evidence-based claims. As such it was helpful, however the reviewers observed in their own rating and commenting behaviour, divergences in how to interpret conceptual reproducibility. Conceptual reproducibility deeply touches on epistemic norms and values inside of a field, and ongoing debates. Our discussion very much centered on how to assess the appropriateness of methods and of claims made based on the evidence produced in the light of unsettled methodological debates in our field. We further observed that there exists a diversity of research designs and methods in our field - what is the risk implied to this arguably productive diversity of methods, if journals take a strong stance on enforcing the use of standard methods? And what is the role of a single article in ensuring conceptual reproducibility, and at what point is a debate to be taken to a wider forum in the community, and if so in what form (i.e., methods sections in journals, controversies addressed at conferences, benchmarking tests in training and education)?

*Limitations*

This explorative study is only a first step in the effort to assess empirically the specific form that threats to the reproducibility of research take in scientometrics. The small, hand-selected sample of publications we reviewed is not representative for the entire body of research published in scientometrics, e.g. in terms of study designs, methods, and data used. We aimed to account for some of the variation we encounter in scientometrics by our high-level categorization of study types. However, due to the smallness of the sample we could not capture the variation within each category of study types. Hence we cannot make conclusive statements about the extent to which reproducibility issues vary by study type, nor whether the distinction made by our categorization is the most relevant one to account for major variations in the type of issues encountered. Finally, we lack an empirical investigation of what types of research best characterize the large majority of studies in our field. These are all topics for future research.

One point of potential concern arising from our explorative review, however, is the initial impression that empirical studies as currently published in our field show weaknesses with regard to the critical reflection on and justification of chosen data sets, methods, and operationalizations relative to the specific research questions asked, omissions to demonstrate the robustness of results against parameter variations, and failures to base claims adequately on the empirical results.

**Conclusion**

The approach we tested here to identify reproducibility issues in scientometrics has been to conduct a multi-reviewer exercise by a team of researchers with a variety of methodological and epistemic backgrounds, who were guided by taxonomies of threats to reproducibility. The application of the taxonomies has been challenging, revealing remaining confusions about concepts of reproducibility and the need for further consolidation or explication of such taxonomies to use them to support such reviewing exercises. That said, the discussions around the assessment of features of studies relative to our at times diverging ideals of reproducible research, have been productive in eliciting open questions regarding their implications in terms of requirements for the publication of scientometric studies. These questions are laid out in this paper and the companion paper on direct reproducibility (Waltman et al. 2018).

For the upcoming STI2018 conference in Leiden, we suggest to discuss some of the questions raised. One of the key questions with regard to conceptual reproducibility is how to operationalize expectations for individual articles with regard to the robustness of their knowledge claims, whether the status of methodological debates in our field allows us to be more prescriptive with regard to the appropriateness of methods, and where such debates are most needed and how they could be best supported.

**References**

Barba, L. A. (2018). *Terminologies for reproducible research*. arXiv:1802.03311.

Chen, X. (1994). The rule of reproducibility and its applications in experiment appraisal. *Synthese*, *99*(1), 87–109.

Fidler, F., Chee, Y. E., Wintle, B. C., Burgman, M. A., McCarthy, M. A., & Gordon, A. (2017). Metaresearch for evaluating reproducibility in ecology and evolution. *BioScience*, *67*(3), 282–289.

Flier, J. S. (2017). Irreproducibility of published bioscience research: Diagnosis, pathogenesis and therapy. *Molecular Metabolism*, *6*(1), 2–9.

Goodman, S. N., Fanelli, D., & Ioannidis, J. P. A. (2016). What does research reproducibility mean? *Science Translational Medicine*, *8*(341), 341ps12.

Open Science Collaboration (2015). Estimating the reproducibility of psychological science. *Science*, *349*(6251), aac4716.

Pashler, H., & Harris, C. R. (2012). Is the replicability crisis overblown? Three arguments examined. *Perspectives on Psychological Science*, *7*(6), 531–536.

Schneider, J. W. (2015). Null hypothesis significance tests. A mix-up of two different theories: The basis for widespread confusion and numerous misinterpretations. *Scientometrics*, *102*(1), 411–432.

Simmons, J. P., Nelson, L. D., & Simonsohn, U. (2011). False-positive psychology: Undisclosed flexibility in data collection and analysis allows presenting anything as significant. *Psychological Science*, *22*(11), 1359–1366.

Waltman, L., Hinze, S., Scharnhorst, A., Schneider, J.W., Velden, T., Exploration of reproducibility issues in scientometric research - Part 1: Direct reproducibility(*submitted*) In: Proceedings STI 2018. Leiden September 12-14, 2018.